\documentclass[amssymb,prd,aps,showpacs,nofootinbib]{revtex4}
\usepackage{latexsym}
\usepackage{amsmath}
\usepackage[T1]{fontenc}
\makeindex
\newcommand{\be}{\begin{equation}}
\newcommand{\ee}{\end{equation}}
\newcommand{\bn}{\begin{eqnarray}}
\newcommand{\en}{\end{eqnarray}}
\newcommand{\bns}{\begin{eqnarray*}}
\newcommand{\ens}{\end{eqnarray*}}

\def\VE*{\vec{E}^{*}}

\def\ba{\begin{eqnarray}}
\def\ea{\end{eqnarray}}

\begin{document}
\hfill Preprint CBPF-NF-037/03
%\onehalfspacing
\title{Charged Tensor Matter Fields and Lorentz Symmetry Violation via Spontaneous Symmetry Breaking}
%\indent
\author{L.P. Colatto}
\email{colatto@cbpf.br}
\affiliation{~CCP, CBPF, Rua Xavier
Sigaud, 150, CEP 22290-180, Rio de Janeiro, RJ, Brasil\\}
\affiliation{Grupo de F\'{\i}sica Te\'orica Jos\'e Leite Lopes,
C.P. 91933, CEP 25685-970, Petr\'opolis,  RJ, Brasil}
\author{A.L.A. Penna}
\email{andrel@fis.unb.br}
\author{W.C. Santos}
\email{wytler@fis.unb.br}
\affiliation{{\small IF, Universidade de
Bras\'\i lia, CEP 70919-970, Bras\'\i lia, DF, Brasil}}


\begin{abstract}
\noindent
We consider a model with a charged vector field along with a Cremmer-Scherk-Kalb-Ramond (CSKR) matter field coupled to a $U(1)$ gauge potential. We obtain a natural Lorentz symmetry violation due to the local $U(1)$ spontaneous symmetry breaking mechanism triggered by the imaginary part of the vector matter. The choice of the unitary gauge leads to the decoupling of the gauge-KR sector from the Higgs-KR sector. The excitation spectrum is carefully analyzed and the physical modes are identified. We propose an identification of the neutral massive spin-$1$ Higgs-like field with the massive $Z'$ boson of the so-called mirror matter models.
\end{abstract}
\date{\today }
\preprint{CBPF-Notas de Física}
\maketitle

\section{Introduction}

Systematic studies of the relations between topological models in $(1+2)D$ and
the phenomenology of planar theories, as high-$T_{c}$ supercondutivity have been taken into account
since the formulation of the Chern-Simons theory \cite{linde,jackiw}. One remarkable characteristic extracted from the dynamics of the high-$T_{c}$ superconductors is the violation of the $P$- and the $T$-symmetries. This fact emphasizes its planar nature.
As a matter of fact, topological models originated from Chern-Simons term are
restricted to describe objects living in $(1+2)D$. This aspect has motivated the study of extensions of planar gauge theories and the mathematical properties underlined, such as the fractional statistics \cite{wilczek}. Relevant extensions are the complex Maxwell-Chern-Simons (MCS$^*$) model in $(1+2)D$ \cite{schon} and, the complex Maxwell-Chern-Simons-Proca theory(MCSP$^{*}$). The planar scenario and the dynamical mass generation of these two models were largely exploited in Ref. \cite{osw}. In this context, the physical investigation of topological dynamical aspects of complex matter vector fields may still be better explored, specially in $(1+3)D$ high-energy physics as well as in condensed matter systems.

The concept of topological models in $(1+3)D$ has been firstly pointed out by Cremmer and Scherk \cite{cs}, and Kalb and Ramond \cite{kr}; we refer to this class of models as the CSKR model. In these works, a topological term in $(1+3)D$ is introduced which is an extension to the well-known Chern-Simons term in $(1+2)D$. This topological CSKR term introduces a direct coupling between a 1-form gauge field and another 2-form gauge field without an effective
contribution to the energy and the momentum of the model; it however gives a mass contribution at tree-level. The CSKR, as a topological model, is another candidate to generate mass without introducing a Higgs scalar field into the Lagrangian. So, gauge  symmetry is preserved and a massive spin-$1$ boson appears. Consequently we can inquire whether a charged spin-$1$ vector boson could be incorporated into the spectrum of the CSKR model. Incidentally, vector-tensor field models have been largely studied, particularly in the context of $N=2$ supersymmetric models \cite{N2}. On another hand, the contraction of a Chern-Simons term with a fixed vector has been proposed in order to build up a Lorentz-violating model to describe astrophysical effects and cosmological new perspectives (possibly from geometrical origin) due to the variation of the universal constants \cite{astro}. Indeed, recent works have suggesting models to study Lorentz violation:
D. Colladay and Kostelecký \cite{colla1,colla2} suggested a general Lorentz-violation extension of the Standard Model including CPT-even and CPT-odd terms in $(1+3)D$. They have obtained that the extension presents a gauge invariance and conserved energy-momentum tensor while covariance under particle rotations and boosts is broken. Coleman and Glashow \cite{cole1,cole2} have also investigated tiny non-invariant terms introduced into the standard model Lagrangian in a perturbative framework. The effects of these perturbations
increase rapidly with the energy for a preferred frame what implies in a Lorentz-violation of the system. The occurrence of the dynamical breaking of the Lorentz symmetry in Abelian vector field models with the  Wess-Zumino interaction have been explored by Andrianov and Soldati \cite{and1,and2}. On the other hand, Carrol, Field and Jackiw \cite{jackiw3} have
demonstrated that ordinary Chern-Simons terms, studies previously in $(1+2)D$, can couple to dual electromagnetic tensor to an fixed and external four-vector. The effects of this Chern-Simons terms in Lorentz and CPT-violation have also been treated by Jackiw and Kostelecký \cite{jackiw4}.

The aim of this work is to study a charged vector-tensor matter field model based on  the complex extension of the CSKR model. We build up a full Lagrangian model where  all the possible invariant terms are included.  Furthermore, we add up a local $U(1)$ symmetry in order to have an interacting charged vector-tensor field model where a gauge field $A_{\mu}$ mediates the interaction. In this model, the 1- and 2-form fields can coexist and interact with each other by means of a topological term in $(1+3)D$. To check the consistency in a quantum field-theoretic sense, we discuss aspects such as causality, unitarity of the excitation spectrum. To this aim, we take into account the local $U(1)$ interaction formulation and we will analyze the vacuum states in the low-energy limit.  

Due to the vector nature of the order parameter of the model, the ground state is identified with a constant four-vector (we call $b_{\mu}$) which implies an anisotropy of the vacuum state as a by-product and naturally induces Lorentz symmetry violation. This vacuum anisotropy has recently received much attention in connection with astrophysical phenomena \cite{astro1}.  We study the role played by the vector $b_{\mu}$ and its consequences to the physical degrees of freedom described in the spectrum of the model. We also explore the possible consistent (no ghost and no tachyon) choices of this vector background.

The outline of our paper is as follows: in Section II, we introduce the full global $U(1)$ vector-tensor matter field model and obtain the equations of motion, the Noether and the topological currents. In Section III, using the hint of accommodating the two kind of currents in a duplet, we compute the propagators, poles and the physical consistency relation obeyed. In Section IV, we switch on an interacting gauge field and introduce a local $U(1)$ symmetry. We study the SSB mechanism and conclude that the potential achieves its minimum for a non-vanishing vacuum expectation value of the charged vector matter field. We adopt to work with the unitary gauge and, in Section V, we study the spectrum and consistency relations for the gauge-KR sector. In Section VI, the Higgs-KR sector is analyzed. Finally, in Section VII, we discuss and comment our results.

\section{The Global $U(1)$ Vector-Tensor Field Model}
\noindent
Based on the CSKR model, we propose to study a full $U(1)$ charged vector-tensor
matter field model \cite{cs,kr} where
we have also included the topological terms. It can be written down as
\begin{eqnarray}
\label{Lag1}
 {\cal L}&=&
\frac{1}{3}G_{\mu\nu\kappa}^{*}G^{\mu\nu\kappa}
-\frac{1}{2}F_{\mu\nu}^{*}F^{\mu\nu}-
(\partial_{\mu}B^{\mu})^{*}(\partial_{\nu}B^{
\nu})+2(\partial_{\mu}H^{\mu\nu})^{*}(\partial^{\rho}H_{\rho\nu})+ \nonumber
\\ \nonumber \\
& & + \alpha^2B_{\mu}^{*}B^{\mu} +\lambda (B_{\mu}^{*}B^{\mu})^2- \beta
^2H_{\mu\nu}^{*}H^{\mu\nu} + m
\epsilon^{\mu\nu\rho\sigma}B_{\mu}^{*}\partial_{\nu}H_{\rho\sigma}
+m\epsilon^{\mu\nu\rho\sigma}B_{\mu}\partial_{\nu}H_{\rho\sigma}^{*},
\end{eqnarray}
where $B^{\mu}$ and $H^{\mu \nu}$ are the matter vector and tensor fields
respectively, $\alpha$ and $\beta$ represent the mass term parameters of the fields,
$\lambda$ represent the self-interacting parameter and
$m$ the topological mass\footnote{We also
adopt the metric $(+,-,-,-)$.}.  The field strengths can be defined by
\begin{eqnarray}
G_{\mu\nu\kappa}&=&\partial_{\mu}H_{\nu\kappa}+
\partial_{\nu}H_{\kappa\mu}+\partial_{\kappa}H_{\mu\nu} ,\nonumber\\
F_{\mu\nu}&=&\partial_{\mu}B_{\nu}-\partial_{\nu}B_{\mu}.
\end{eqnarray} 
We observe that the
topological term is a mixed one formed by
$B_{\mu}$ and $H_{\mu\nu}$ in $4D$ as the term studied in the Ref. \cite{cs,kr}.
Consequently, $m$ is regarded as a topological mass. We also consider a potential
term that defines the quadratic mass parameters $\alpha^2$ and
$\beta^2$. The conserved matter current $J^{\mu}$ stemming from the global $U(1)$ symmetry
is given by,
\begin{eqnarray}
J^{\mu}&=& i (B_{\nu}F^{\mu\nu *}-
B_{\nu}^*F^{\mu\nu})- i(H_{\nu\kappa}^{*}G^{\mu\nu\kappa} -
H_{\nu\kappa}G^{\mu\nu\kappa *}) +\nonumber \\ \nonumber
&&+ im
\epsilon^{\mu\nu\kappa\lambda}(B_{\nu}^*H_{\kappa\lambda}-
B_{\nu}H_{\kappa\lambda}^*)
+ i[B^{\mu}(\partial_{\nu}B^{\nu})^*-B^{\mu
*}(\partial_{\nu}B^{\nu})]+ \nonumber \\
&&-i
[(\partial^{\rho}H_{\rho\nu})^*H^{\mu\nu}-(\partial^{\rho}H_{\rho\nu})H^{\mu\nu
*}]. \label{current}
\end{eqnarray}
The coupled Euler-Lagrange equations are
\begin{eqnarray}
\Box B^{\nu}&=& -\alpha^2 B^{\nu}- \lambda (B_{\mu}^* B^{\mu})B^{\nu}-
m\epsilon^{\nu\kappa\lambda\rho}\partial_{\kappa}H_{\lambda\rho},\nonumber\\
\Box H^{\nu\kappa}&=&-\beta^2H^{\nu\kappa}+
m\epsilon^{\nu\kappa\lambda\rho}\partial_{\lambda}B_{\rho}.
\end{eqnarray}
According to the symmetry of the Lagrangian (\ref{Lag1}), we
notice the occurrence of a $B^4$-interaction term that determines a
anti-symmetrized identically conserved topological current of type,
\begin{equation}
\label{topological}
J^{\mu\nu}=\frac{1}{2}\epsilon^{\mu\nu\kappa\lambda}\partial_{\kappa}B_{\lambda}.
\end{equation}
Its associated topological-vector-charge gives rise to ``vector solitons''
solution whose value may be regarded as a quantum number.  Indeed, this topological current  induces directly the nonlinear behavior on the vector-matter field sector.
The investigation of the nonlinear dynamic and the non-trivial configuration of the fields with anisotropic energy in $(1+2)D$ will be explored in a forthcoming work.

\section{The Spectrum Analysis}

In order to verify the physical spectrum, we rearrange the Lagrangian (\ref{Lag1})
in a linearized form, or
 \begin{equation}
 {\cal L}={\bf V}^t {\cal O}{\bf V},
 \end{equation}
 where $\cal O$ is a unitary wave operator and we represent $\bf V$ as a
 vector-tensor duplet or,
 \begin{equation}
 {\bf V}= \begin{pmatrix}
 B_{\mu}\cr \cr
 H_{\mu\nu}
 \end{pmatrix},
 \end{equation}
To obtain the propagators by means of the usual mechanism, we
take the ${\cal O}^{-1}$ using the usual product algebra of the ordinary
longitudinal, transverse and spin projector operators, which respectively are
$\omega _{\mu\nu}$,
$\theta _{\mu\nu}$ and
$s_{\mu\nu\lambda}={\epsilon^{\gamma}}_{\mu\nu\lambda}\partial _{\gamma}$.
In addition, we also have the anti-symmetric longitudinal and transverse
spin four-indexed projector operators written down as,
\begin{equation}
\theta_{\mu\nu,\lambda\rho}=\frac{1}{2}(\theta_{\mu\lambda}\theta_{\nu\rho}-
\theta_{\mu\rho}\theta_{\nu\lambda} ) ,\hspace{1cm}
\omega_{\mu\nu,\lambda\rho}=\frac{1}{2}(\eta_{\mu\lambda}\omega_{\nu\rho}-
\eta_{\mu\rho}\omega_{\nu\lambda}),\hspace{1cm}
\eta_{\mu\nu,\lambda\rho}=\frac{1}{2}(\eta_{\mu\lambda}\eta_{\nu\rho}-
\eta_{\mu\rho}\eta_{\nu\lambda} ),
\end{equation}
which implies a closed algebra, such that
\begin{center}
\begin{tabular}{|l|c|c|c|}
\hline &$\hspace{0.5cm}{\theta^{\alpha\beta}}_{,\lambda\rho} \hspace{0.5cm}$ & $\hspace{0.5cm}{\omega^{\alpha\beta}}_{,\lambda\rho}\hspace{0.5cm}$ &  $\hspace{0.5cm}{s^{\alpha\beta}}_{\lambda}$\hspace{0.5cm} \\
\cline{1-4} \hline $\hspace{0.5cm} \theta_{\mu\nu,\alpha\beta}$\hspace{0.5cm} &$\hspace{0.5cm}\theta_{\mu\nu,\lambda\rho}\hspace{0.5cm}$ & $\hspace{0.5cm}0\hspace{0.5cm}$ &$\hspace{0.5cm}s_{\mu\nu\lambda}\hspace{0.5cm}$ \\
\cline{1-4} \hline $\hspace{0.5cm} \omega_{\mu\nu,\alpha\beta}$&$\hspace{0.5cm}0\hspace{0.5cm}$ &$\hspace{0.5cm}\omega_{\mu\nu,\lambda\rho}\hspace{0.5cm}$ &$\hspace{0.5cm}0\hspace{0.5cm}$ \\
\cline{1-4} \hline $\hspace{0.5cm} {s_{\mu\alpha\beta}}$ &$\hspace{0.5cm}{s_{\mu\lambda\rho}}\hspace{0.5cm}$ &$\hspace{0.5cm}0\hspace{0.5cm}$ &$\hspace{0.5cm}-\Box\theta_{\mu\lambda}\hspace{0.5cm}$ \\
\cline{1-4} \hline
\end{tabular}\, ,
\end{center}
where we have obtained that $
\eta_{\mu\nu,\lambda\rho}=\omega_{\mu\nu,\lambda\rho}+
\theta_{\mu\nu,\lambda\rho}$.
We obtain the propagators, in the momentum
space, which can be written down as
\begin{eqnarray}
<B_{\mu}^{*},B_{\nu}>&=&\frac i{(k^2-\alpha^2)}\omega_{\mu\nu}+
\frac{i(k^2-\beta^2)}{(k^2-\mu^2_{+})(k^2-\mu^2_{-})}
\theta_{\mu\nu},\nonumber \\
<B_{\mu}^*,H_{\nu\lambda}>&=& <H_{\mu\nu}^*,B_{\lambda}> =
\frac{im}{(k^2-\mu^2_{+})(k^2-\mu^2_{-})}s_{\mu\nu\lambda},\nonumber \\
<H_{\mu\nu}^{*},H_{\lambda\rho}>&=&\frac i{(k^2-\alpha^2)}
\omega_{\mu\nu,\lambda\rho}+\frac{i(k^2-\beta^2)}
{(k^2-\mu^2_{+})(k^2-\mu^2_{-})}\theta_{\mu\nu,\lambda\rho},\nonumber\\
\end{eqnarray}
where,
\begin{equation}
\label{mu} \mu_{\pm}^{2}= \frac{\alpha^2+\beta^2+2m^2\pm
\sqrt{(\alpha^2+\beta^2+2m^2)^2-4\alpha^2\beta^2}}{2},
\end{equation}
which can be easily verified to be real and positive. As a consequence, we observe
that the poles $k^2=\alpha^2$, $k^2=\beta^2$,
$k^2=\mu^2_{+}$ and $k^2=\mu^2_{-}$ indicate the absence of tachyon states. Another point is the positivity of the norm of the states verified from the analysis of the residues of the propagators obtained. To do that, we take the transition amplitudes considering the doublet "vector-tensor current",  which can be written down as,
\begin{equation}
{\bf J}= \left( \begin{array}{c}
J_{\mu} \\ J_{\mu\nu} \end{array} \right),
\end{equation}
where $J_{\mu}$ is the usual Noether current, and $J_{\mu\nu}$ is the current given in the eq. (\ref{current}) and (\ref{topological}) which is conserved by definition.

We observe that there are two
dynamical physical poles, both describing massive particles
specified by $\mu_{+}^2$ and $\mu_{-}^2$. The
transverse topological sectors are non-dynamical. We obtained that the propagators $<B_{\mu}^{*},B_{\nu}>$ 
and $<H_{\mu\nu}^{*},H_{\lambda\rho}>$  have the same poles and consequently the same particles. The crossing ones have no dynamics. We can see an order in the spectrum of the model which obeys the relation
\begin{equation}
\mu_{+}>\beta >\alpha > \mu_{-},
\end{equation}
resulting in a consistent physical model. We observe that to perform the analysis of the degrees of freedom in $1+3$ dimensions it was necessary to take the antisymmetric topological  current $J_{\mu\nu}$ given in (\ref{topological}) to complete a doublet with the usual vector one ($J_{\mu}$). Indeed, the topological current has induced directly the nonlinear
behavior of the vector-matter field sector.

\section{The Local $U(1)$ Theory and SSB}

To introduce interactions into the model representing by  the matter Lagrangian (\ref{Lag1}), we take as local the symmetry phase, as the usual method, or
\begin{equation}
\begin{matrix} B_{\mu}^{'} = e^{- i\Lambda(x)} B_{\mu},&\, \,
\text{and} \,\, &
  H_{\mu\nu}^{'} =e^{-i\Lambda(x)} H_{\mu\nu} \end{matrix}.
\end{equation}
In this way the symmetries are restored, introducing the invariant Lagrangian written
as,
\begin{eqnarray}
\label{ultimo} {\cal L}_{int}&=&\frac{1}{3}{\cal
G}_{\mu\nu\kappa}^{*}{\cal G}^{\mu\nu\kappa}-\frac{1}{2}{\cal
F}_{\mu\nu}^{*}{\cal
F}^{\mu\nu}-\frac{1}{4}f^{\mu\nu}f_{\mu\nu}-(D_{\mu}B^{\mu})^{*}(D_{\nu}B^{\nu})+2(D_{\mu}H^{\mu\nu})^{*}
(D^{\rho}H_{\rho\nu})+ \nonumber \\ \nonumber \\
& &+ m
\epsilon^{\mu\nu\kappa\lambda}B_{\mu}^{*}D_{\nu}H_{\kappa\lambda}
+m\epsilon^{\mu\nu\kappa\lambda}B_{\mu}(D_{\nu}H_{\kappa\lambda})^{*}+
\alpha^2B_{\mu}^{*}B^{\mu}+ \lambda (B_{\mu}^{*}B^{\mu})^2 - \beta ^2H_{\mu\nu}^{*}H^{\mu\nu},
\end{eqnarray}
where $D_{\mu}=\partial_{\mu}+ieA_{\mu}$ is the covariant
derivative. The above Lagrangian describes an interaction model between the
matter fields $B_{\mu}$, $H_{\mu\nu}$ and the gauge field
$A_{\mu}$, where we define
\begin{gather}
{\cal G}_{\mu\nu\kappa}=D_{\mu}H_{\nu\kappa}+D_{\nu}H_{\kappa\mu}+
D_{\kappa}H_{\mu\nu},\hspace{1cm} {\cal
F}_{\mu\nu}=D_{\mu}B_{\nu}-D_{\nu}B_{\mu},\hspace{1cm}\mbox{and}\hspace{1cm}
f_{\mu\nu}=\partial_{\mu}A_{\nu}-\partial_{\nu}A_{\mu}.
\end{gather}
The new conserved current can be written down as,
\begin{eqnarray}
\partial_{\mu}f^{\mu\nu}=\cal J^{\nu}&=& i (B_{\nu}{\cal F}^{\mu\nu *}-
B_{\nu}^*{\cal F}^{\mu\nu})- i(H_{\nu\kappa}^{*}{\cal
G}^{\mu\nu\kappa} - H_{\nu\kappa}{\cal G}^{\mu\nu\kappa *})+\nonumber \\
\nonumber \\& &+ i m
\epsilon^{\mu\nu\kappa\lambda}(B_{\nu}^*H_{\kappa\lambda}-
B_{\nu}H_{\kappa\lambda}^*) +i[B^{\mu}(D_{\nu}B^{\nu})^*-B^{\mu
*}(D_{\nu}B^{\nu})]+\nonumber
\\ \nonumber \\& &-i
[(D^{\rho}H_{\rho\nu})^*H^{\mu\nu}-(D^{\rho}H_{\rho\nu})H^{\mu\nu
*}],
\end{eqnarray}
where ${\cal J}^{\nu}$ is the covariant matter current.

As we have seen, the covariant interacting vector-tensor model described by the Lagrangian (\ref{ultimo}) introduces a $U(1)$ gauge field, $A_{\mu}$. To explore the behavior of these fields at low-energy phenomenology, we take the spontaneous breaking of gauge symmetry mechanism. As the model carries vector and tensor fields as matter degrees of freedom, the discussion of the SSB becomes subtle. The quartic self-interacting non-linear term of the matter field $B_{\mu}$ in the Lagrangian (\ref{ultimo}) could play a role similar to the Higgs field, but with a vector nature.  The $\lambda (B_{\mu}^{*}B^{\mu})^2$ does not spoil the invariance of the Lagrangian under the group of the local $U(1)$ transformations. So, the condition of a minimum of the energy ($E$) can be obtained taking the minimum of the potential energy ($V$), or
\begin{equation}
\label{equa}
\frac{dE}{dB_{\mu}}=\frac{dV}{dB_{\mu}}=\alpha^2B^{*}_{\mu}+
2\lambda(B_{\nu}^{*}B^{\nu})B_{\mu}^{*}=0,
\end{equation}
where it is analogous to the $B_{\mu}^*$ term, recalling that $\alpha^2$ is a mass parameter. In this case,
the situation where $\alpha^2<0$,  and $\lambda >0$, introduces a
non-trivial vacuum, and it follows that the energy is a minimum
at,
\begin{eqnarray}
& &B_{\mu}^{*}B^{\mu}=b_{\mu}b^{\mu}=b^2=-\frac{\alpha^2}{2\lambda}u^2,
\label{bmu}
\end{eqnarray}
where we observe that, in this case, we require that $b^2$ be a constant 4-vector parameter
such that $-\frac{\alpha^2}{2\lambda}>0$. In fact, we observe that the VEV for the
field $B_{\mu}$ is given by,
\begin{eqnarray}
\langle 0 |B_{\mu} | 0 \rangle
=b_{\mu}=\sqrt{\frac{-\alpha^2}{2\lambda}}u_{\mu},
\label{rad}
\end{eqnarray}
where $u_{\mu}$ is a unitary vector which lying in a fixed direction in the space-time. In turn, it breaks the Lorentz symmetry, and due to this arbitrariness we have to choose amongst the possible types of vector:
$u^{2}=1$ (time-like), $u^{2}=-1$ (space-like) or $u^{2}=0$ (light-like), analogous to the case studied in the Ref. \cite{belich}.
As we are interested to deal with non-trivial configurations of the fields, we exclude the light-like possibility.  Consequently, we reach a non-trivial vacuum solution for an energy $E$ which breaks spontaneously the $U(1)$ local symmetry, and also violates the Lorentz symmetry. We emphasize that the Lorentz violation came along as by-product effect of the internal symmetry breaking. The Lorentz violation has received much attention due to possible astrophysical and condensed matter effects \cite{astro,cond}, which deserves a more deep analysis.  In this work, we are going to verify the mass spectrum of this model.  To this purpose, we begin by observing that the system under consideration has an infinite set of vacuum states, corresponding to points on a circle of radius given by the Eq. (\ref{rad}) posed on the complex plane of the field $B_{\mu}$.  So, we can decompose the complex fields into components and we shift the field $B_{\mu}$  along the real axis (analogous to the Higgs mechanism). So we have
\begin{eqnarray}
\label{B} B_{\mu} \rightarrow B_{\mu}+b_{\mu}= X_{\mu}+iY_{\mu}
+b_{\mu}, \hspace{1cm}\mbox{and} \hspace{1cm}  H_{\mu\nu}
\rightarrow P_{\mu\nu}+iQ_{\mu\nu},
\label{bandh}
\end{eqnarray}
so, we can express the potential term as,
\begin{equation}
V=\lambda(B_{\mu}^{*}B^{\mu}-b^2)^2-\lambda
b^4-\beta^2H_{\mu\nu}^{*}H^{\mu\nu},
\end{equation}
which are substituted into the Lagrangian (\ref{ultimo}), whose expansion we find,
\begin{eqnarray}
\label{maracatu} {\cal
L}_{broken}&=&\frac{1}{3}P_{\mu\nu\kappa}P^{\mu\nu\kappa}+\frac{1}{3}Q_{\mu\nu\kappa}Q^{\mu\nu\kappa}-\frac{1}{2}X_{\mu\nu}X^{\mu\nu}-\frac{1}{2}Y_{\mu\nu}Y^{\mu\nu}
-(\partial_{\mu}X^{\mu})^2-(\partial_{\mu}Y^{\mu})^2+2(\partial_{\mu}P^{\mu\nu})(\partial^{\rho}P_{\rho\nu})\nonumber
\\&
&+2(\partial_{\mu}Q^{\mu\nu})(\partial^{\rho}Q_{\rho\nu})-e^2b^2A_{\mu}A^{\mu}-\frac{1}{4}f_{\mu\nu}f^{\mu\nu}+2e(b^{\mu}A^{\nu})(\partial_{\mu}Y_{\nu})-2e(A^{\mu}b^{\nu})(\partial_{\mu}Y_{\nu})
+2e(\partial_{\mu}Y^{\mu})(A_{\nu}b^{\nu})\nonumber \\
&
&+2m\epsilon^{\mu\nu\kappa\lambda}X_{\mu}\partial_{\nu}P_{\kappa\lambda}-2em\epsilon^{\mu\nu\kappa\lambda}b_{\mu}A_{\nu}Q_{\kappa\lambda}
+4\lambda
b_{\mu}b_{\nu}X^{\mu}X^{\nu}-\beta^2P_{\mu\nu}P^{\mu\nu}-
\beta^2Q_{\mu\nu}Q^{\mu\nu} \nonumber \\ &&+ \mbox{higher order
terms},
\end{eqnarray}
where  $X_{\mu\nu}$, $Y_{\mu\nu}$, $P_{\mu\nu\kappa}$ and $Q_{\mu\nu\kappa}$, are the field-strengths of their respective real components of 1- and 2-form fields written in the definitions Eq. (\ref{bandh}).  We observe  that the Lagrangian (\ref{maracatu}) is non-diagonal, what become a subtle computation. The terms
$2e(b^{\mu}A^{\nu})(\partial_{\mu}Y_{\nu})$,
$-2e(A^{\mu}b^{\nu})(\partial_{\mu}Y_{\nu})$ and
$2e(\partial_{\mu}Y^{\mu})(A_{\nu}b^{\nu})$ can be absorbed by carrying out the following field re-definitions,
\begin{eqnarray}
A_{\mu}\rightarrow
A_{\mu}-q_{\nu}(\partial_{\mu}Y^{\nu})+q_{\nu}(\partial^{\nu}Y_{\mu})+q_{\mu}(\partial_{\nu}Y^{\nu}), \hspace{0.5cm}\mbox{and}\hspace{0.5cm}
f_{\mu\nu} \rightarrow f_{\mu\nu}+\gamma
Y_{\mu\nu}+\Sigma_{\mu\nu}(\partial_{\alpha}Y^{\alpha}),
\label{transf}
\end{eqnarray}
where $q_{\nu}$, $\gamma$ and $\Sigma_{\mu\nu}$ are operators that can be easily found by manipulating (\ref{transf}) and (\ref{maracatu}), which can be define as
\begin{eqnarray}
q^{\nu}=\frac{b^{\nu}}{eb^2},\hspace{0.7cm}\gamma=q^{\alpha}\partial_{\alpha}, \hspace{0.7cm}\mbox{and} \hspace{0.7cm}
\Sigma_{\mu\nu}=q_{\mu}\partial_{\nu}-q_{\nu}\partial_{\mu}.
\end{eqnarray}
So, the resulting  Lagrangian (\ref{maracatu}) is
non-gauge-invariant because the Lorentz symmetry is broken. It breaks translation due to the presence of the $\gamma$ operator, and it breaks rotation by virtue of the
$\Sigma_{\mu\nu}$ operator present in the new definitions in Eq. (\ref{transf}). Nevertheless, we can eliminate the $Y_{\mu}$ field  by means of a gauge choice, picking 
a particular gauge parameter in the $U(1)$ phase transformation; so,
\begin{equation}
\label{gaugeunitario}
\begin{matrix}
X^{'}_{\mu}=X_{\mu}-\Lambda Y_{\mu}, \cr\cr
Y^{'}_{\mu}=Y_{\mu}-\Lambda X_{\mu}+\Lambda b_{\mu},
\end{matrix}
\end{equation}
where $\Lambda$ is an arbitrary gauge parameter. In fact, we can gauge away the $Y_{\mu}$ field choosing a particular gauge, bearing in mind the unitarity condition on the particle spectrum.  Then $A_{\mu}$ field acquires mass due to the presence of the scalar field-parameter $\Lambda=\Phi$ in its longitudinal mode. This describes the associated Higgs-Mechanism for a complex vector. It can be seen through the following re-defined transformations,
\begin{equation}
A_{\mu}\rightarrow A_{\mu}-
\partial_{\mu} \Phi,\hspace{1cm}
f_{\mu\nu}^{'}=f_{\mu\nu}.
\end{equation}
Then, the Lagrangian (\ref{maracatu}) can be rewritten in the
absence of the interaction terms as,
\begin{eqnarray}
\label{maracatu2} {\cal
L}_{broken}&=&\frac{1}{3}P_{\mu\nu\kappa}P^{\mu\nu\kappa}+\frac{1}{3}Q_{\mu\nu\kappa}Q^{\mu\nu\kappa}-\frac{1}{2}X_{\mu\nu}X^{\mu\nu}
-(\partial_{\mu}X^{\mu})^2+2(\partial_{\mu}P^{\mu\nu})(\partial^{\rho}P_{\rho\nu})+2(\partial_{\mu}Q^{\mu\nu})(\partial^{\rho}Q_{\rho\nu})+
\nonumber \\&&-\,e^2b^2A_{\mu}A^{\mu}
-\frac{1}{4}f_{\mu\nu}f^{\mu\nu}
+2m\epsilon^{\mu\nu\kappa\lambda}X_{\mu}\partial_{\nu}P_{\kappa\lambda}-2em\epsilon^{\mu\nu\kappa\lambda}b_{\mu}A_{\nu}Q_{\kappa\lambda}
+4\lambda
b_{\mu}b_{\nu}X^{\mu}X^{\nu}+ \nonumber \\
&&-\beta^2P_{\mu\nu}P^{\mu\nu}-\beta^2Q_{\mu\nu}Q^{\mu\nu}.
\end{eqnarray}
In order to extract the physical content of the Lagrangian
(\ref{maracatu2}) one can split it in two sectors,
\begin{eqnarray}
 \label{maracatu4}{\cal
L}_{gauge-KR}&=&\frac{1}{3}Q_{\mu\nu\kappa}Q^{\mu\nu\kappa}-\frac{1}{4}f_{\mu\nu}f^{\mu\nu}+2(\partial_{\mu}Q^{\mu\nu})(\partial^{\rho}Q_{\rho\nu})
-e^2b^2A_{\mu}A^{\mu}-2em\epsilon^{\mu\nu\kappa\lambda}b_{\mu}A_{\nu}Q_{\kappa\lambda}+ \nonumber \\
&&-\beta^2Q_{\mu\nu}Q^{\mu\nu},\\
\nonumber \\
 {\cal
L}_{Higgs-KR}&=&\frac{1}{3}P_{\mu\nu\kappa}P^{\mu\nu\kappa}-\frac{1}{2}X_{\mu\nu}X^{\mu\nu}
-(\partial_{\mu}X^{\mu})^2+2(\partial_{\mu}P^{\mu\nu})(\partial^{\rho}P_{\rho\nu})
+2m\epsilon^{\mu\nu\kappa\lambda}X_{\mu}\partial_{\nu}P_{\kappa\lambda}\nonumber
\\& & +4\lambda
b_{\mu}b_{\nu}X^{\mu}X^{\nu}-\beta^2P_{\mu\nu}P^{\mu\nu},
\label{maracatu5}
\end{eqnarray}
where we can observe from the Lagrangian (\ref{maracatu4}) that the gauge field only interacts, via topological term, with the imaginary part of the tensor KR field. We can also observe that the mass of the $A_{\mu}$ field depends on the vector $b_{\mu}$, the broken parameter, and on the topological mass as well. On the other hand, the Lagrangian (\ref{maracatu5}) indicates that the vector $b^{\mu}$ contributes to the mass of the real part of the original neutral meson field. We emphasize that the appearance of this new boson field does not prescribe any new symmetry group in the model. To verify the consistency, we are going to compute the spectral analysis separately.

\section{The Spectrum of the Gauge-KR Sector}

A remarkable feature of the Lagrangian (\ref{maracatu4}) is that
it contains a massive gauge vector field (Proca) that interacts with a 2-form KR field. An  analysis of the physical degrees of freedom requires to deal with the unitary gauge (\ref{gaugeunitario}). Then, taking that the fields are well-behaved asymptotically, we can rearrange the Lagrangian considering a mixed doublet defined as ${\bf
U}^t=(A_{\mu}, Q_{\mu\nu})$. So, the Lagrangian  ${\cal L}_{gauge-KR}={\bf U}^t{\cal O}{\bf U}$ where ${\cal O}$ can be easily written down from the Lagrangian (\ref{maracatu4}).
From the inverse operator, ${\cal O}^{-1}$ we can obtain the propagators, in the momentum space, as
\begin{eqnarray}
<A_{\mu},A_{\nu}>&=&\frac {i}{(k^2+e^2b^2)}\Lambda_{\mu\nu}+
\frac{i(k^2-\beta^2)}{(k^2-\tau^2_{+})(k^2-\tau^2_{-})}
\Omega_{\mu\nu}, \nonumber\\
<A_{\mu},Q_{\nu\lambda}>&=&<Q_{\mu\nu},A_{\lambda}> \;= \frac{i
m}{(k^2-\tau^2_{+})(k^2-\tau^2_{-})}\Pi_{\mu\nu\lambda},
\\
<Q_{\mu\nu},Q_{\lambda\rho}>&=&\frac i{(k^2+e^2b^2)}
\Lambda_{\mu\nu,\lambda\rho}+\frac{i(k^2-\beta^2)}
{(k^2-\tau^2_{+})(k^2-\tau^2_{-})}\Omega_{\mu\nu,\lambda\rho},
\nonumber
\end{eqnarray}
in this above expressions we have used the longitudinal,
$\Lambda_{\mu\nu}$,  and transverse, $\Omega_{\mu\nu}$ and
$\Pi_{\mu\nu\lambda}$, given by,
\begin{equation}
\Lambda_{\mu\nu}=\frac{b_{\mu}b_{\nu}}{b^2},
\hspace{1cm}
\Omega_{\mu\nu}=\eta_{\mu\nu}-\frac{b_{\mu}b_{\nu}}{b^2}\hspace{1cm}\mbox{and}
\hspace{1cm}
\Pi_{\mu\nu\lambda}={\epsilon^{\gamma}}_{\mu\nu\lambda}b_{\gamma}.
\end{equation}
whose multiplicative table looks as below:
\begin{center}
\begin{tabular}{|l|c|c|c|}
\hline &$\hspace{0.5cm}{\Omega^{\alpha}}_{\nu} \hspace{0.5cm}$ & $\hspace{0.5cm}{\Lambda^{\alpha}}_{\nu}\hspace{0.5cm}$ &  $\hspace{0.5cm}{\Pi^{\alpha}}_{\nu\lambda}$\hspace{0.5cm} \\
\cline{1-4} \hline $\hspace{0.5cm} \Omega_{\mu\alpha}$\hspace{0.5cm} &$\hspace{0.5cm}\Omega_{\mu\nu}\hspace{0.5cm}$ & $\hspace{0.5cm}0\hspace{0.5cm}$ &$\hspace{0.5cm}\Pi_{\mu\nu\lambda}\hspace{0.5cm}$ \\
\cline{1-4} \hline $\hspace{0.5cm} \Lambda_{\mu\alpha}$&$\hspace{0.5cm}0\hspace{0.5cm}$ &$\hspace{0.5cm}\Lambda_{\mu\nu}\hspace{0.5cm}$ &$\hspace{0.5cm}0\hspace{0.5cm}$ \\
\cline{1-4} \hline $\hspace{0.5cm} {\Pi_{\mu\alpha}}^{\lambda}$ &$\hspace{0.5cm}{\Pi_{\mu\nu}}^{\lambda}\hspace{0.5cm}$ &$\hspace{0.5cm}0\hspace{0.5cm}$ &$\hspace{0.5cm}-b^2\Omega_{\mu\nu}\hspace{0.5cm}$ \\
\cline{1-4} \hline
\end{tabular}
\end{center}
We can also define the anti-symmetrized longitudinal and
transverse operators as,
\begin{equation}
\Omega_{\mu\nu,\lambda\rho}=\frac{1}{2}(\Omega_{\mu\lambda}\Omega_{\nu\rho}-
\Omega_{\mu\rho}\Omega_{\nu\lambda} ) ,\hspace{1cm}
\Lambda_{\mu\nu,\lambda\rho}=\frac{1}{2}(\eta_{\mu\lambda}\Lambda_{\nu\rho}-
\eta_{\mu\rho}\Lambda_{\nu\lambda}),\hspace{1cm}
\eta_{\mu\nu,\lambda\rho}=\frac{1}{2}(\eta_{\mu\lambda}\eta_{\nu\rho}-
\eta_{\mu\rho}\eta_{\nu\lambda} ),
\end{equation}
which implies  the closed product algebra:
\begin{center}
\begin{tabular}{|l|c|c|c|}
\hline &$\hspace{0.5cm}{\Omega^{\alpha\beta}}_{,\lambda\rho} \hspace{0.5cm}$ & $\hspace{0.5cm}{\Lambda^{\alpha\beta}}_{,\lambda\rho}\hspace{0.5cm}$ &  $\hspace{0.5cm}{\Pi^{\alpha\beta}}_{\lambda}$\hspace{0.5cm} \\
\cline{1-4} \hline $\hspace{0.5cm} \Omega_{\mu\nu,\alpha\beta}$\hspace{0.5cm} &$\hspace{0.5cm}\Omega_{\mu\nu,\lambda\rho}\hspace{0.5cm}$ & $\hspace{0.5cm}0\hspace{0.5cm}$ &$\hspace{0.5cm}\Pi_{\mu\nu\lambda}\hspace{0.5cm}$ \\
\cline{1-4} \hline $\hspace{0.5cm} \Lambda_{\mu\nu,\alpha\beta}$&$\hspace{0.5cm}0\hspace{0.5cm}$ &$\hspace{0.5cm}\Lambda_{\mu\nu,\lambda\rho}\hspace{0.5cm}$ &$\hspace{0.5cm}0\hspace{0.5cm}$ \\
\cline{1-4} \hline $\hspace{0.5cm} {\Pi_{\mu\alpha\beta}}$ &$\hspace{0.5cm}{\Pi_{\mu\lambda\rho}}\hspace{0.5cm}$ &$\hspace{0.5cm}0\hspace{0.5cm}$ &$\hspace{0.5cm}-b^2\Omega_{\mu\lambda}\hspace{0.5cm}$ \\
\cline{1-4} \hline
\end{tabular}\, ,
\end{center}
where in addition we can observe that $
 \eta_{\mu\nu,\lambda\rho}=\Lambda_{\mu\nu,\lambda\rho}+
\Omega_{\mu\nu,\lambda\rho}$.
The mass values, $\tau^2_{\pm}$, are easily obtained as,
\begin{eqnarray}
\tau^2_{\pm}&=& \frac{e^2b^2-\beta^2\pm \sqrt{(\beta^2+e^2b^2)^2+
8m^2e^2b^2}}{2}\hspace{1.3cm}\mbox{to the case where $u^2=1$}\nonumber \\
\\
\tau^2_{\pm}&=& \frac{-e^2b^2-\beta^2\pm \sqrt{(\beta^2-e^2b^2)^2-
8m^2e^2b^2}}{2}\hspace{1cm}\mbox{to the case where $u^2=-1$}\nonumber.
\end{eqnarray}
For the case $u^2=1$ (time-like condition), only $\tau_{+}$ defines a
physical excitation. $\tau_{-}$ is a tachyonic excitation.
On the other hand, to $u^2=-1$ (space-like condition), both
$\tau_{+}$ and $\tau_{-}$ are physical excitations for the
restrict values of the topological mass,
\begin{equation}
m<\frac{\beta^2-e^2b^2}{\sqrt{8}eb},
\end{equation}
where $b=|b_{\mu}|$. Observe that $\beta$ cannot be zero, and $\beta>e|b|$, resulting that the Lorentz symmetry breaking  implies  an anisotropy of the space-time is realized as a mass generation on the gauge field, similar to the Higgs mechanism whose consistency is ensured by the presence of a mass term for the 2-form $H_{\mu\nu}$ field.

\section{The Spectrum of the Higgs-KR Sector}

Now, we are going to verify the physical spectrum of the Higgs-KR sector. In a similar way as in the previous case, we are also going to analyze the consequences of the breaking on the spectrum of the Lagrangian (\ref{maracatu5}). We can verify that due to the presence of an anisotropic space-time, the $X_{\mu}$ neutral vector field is deformed, which implies  non-defined propagator poles. Therefore, the physical states can be only obtained by considering the projections of the $X^{\mu}$ field along, and perpendicular to the $b_{\mu}$ vector.
Then defining the $W_{\mu}$ and
$Z_{\mu}$ as the parallel and transverse projections of the $X_{\mu}$ field respectively, or
\begin{eqnarray}
\label{transverso}
W_{\mu}&=&\frac{b_{\alpha}X^{\alpha}}{b^2}\, b_{\mu}\\
Z_{\mu}&=&X_{\mu}-\frac{b_{\alpha}X^{\alpha}}{b^2}\,b_{\mu}.
\end{eqnarray}
We can perform rotations at each point in deformed-space where the
doublet ${\bf U}^t=(X_{\mu}, P_{\mu\nu})$ can be taken as,
\begin{eqnarray}
{\bf U}^t&=&(X_{\mu}, P_{\mu\nu}) \Rightarrow \{{\bf U}{(W)}^t=(W_{\mu},
P_{\mu\nu})_{longitudinal}|\, W_{\mu}=\frac{b_{\alpha}X^{\alpha}}{b^2}\, b_{\mu}, \,\,Z_{\mu}=0\}\\
{\bf U}^t&=&(X_{\mu}, P_{\mu\nu}) \Rightarrow \{{\bf
U}{(Z)}^t=(Z_{\mu}, P_{\mu\nu})_{perpendicular}|\,
Z_{\mu}=X_{\mu}-\frac{b_{\alpha}X^{\alpha}}{b^2}\, b_{\mu},
\,\,W_{\mu}=0\}
\end{eqnarray}
The doublets ${\bf U}(W)$ and ${\bf U}(Z)$ are orthogonal and must be chosen
non-simultaneously. Then we can now to rewrite the Lagrangian
(\ref{maracatu5}) for each one of the above situations, which for the ${\bf U}(W)$ case as yields as
\begin{equation}
\label{Superficie1} {\cal
L}(W)=-P_{\mu\nu}(\eta^{\mu\nu,\lambda\rho}\Box)
P_{\lambda\rho}-W_{\mu}(\eta^{\mu\nu}\Box) W_{\nu}+4\lambda
b^2W_{\mu}W^{\mu}-\beta^2P_{\mu\nu}P^{\mu\nu}+m
\epsilon^{\mu\nu\rho\sigma}W_{\mu}\partial_{\nu}P_{\rho\sigma}
-m\epsilon^{\mu\nu\rho\sigma}P_{\rho\sigma}\partial_{\nu}W_{\mu},
\end{equation}
hence, we can note that by substituting the redefinition of the $W_{\mu}$ field (\ref{transverso}) into
(\ref{maracatu5}) the cross massive term $4\lambda
b_{\mu}b_{\nu}X^{\mu}X^{\nu}$ is transformed to $4\lambda
b^2W_{\mu}W^{\mu}$ and so given a mass term form. On the other hand, we can choose
the $X_{\mu}$ field perpendicular to $b_{\mu}$, where one can
write (\ref{maracatu5}) in the ${\bf U}(Z)$ case. We have that
\begin{equation}
\label{Superficie2}
{\cal L}{(Z)}=-P_{\mu\nu}(\eta^{\mu\nu,\lambda\rho}\Box)
P_{\lambda\rho}+Z_{\mu}(\eta^{\mu\nu}\Box)
Z_{\nu}-\beta^2P_{\mu\nu}P^{\mu\nu}+m
\epsilon^{\mu\nu\rho\sigma}Z_{\mu}\partial_{\nu}P_{\rho\sigma}
-m\epsilon^{\mu\nu\rho\sigma}P_{\rho\sigma}\partial_{\nu}Z_{\mu} ,
\end{equation}
where the mass term $4\lambda b_{\mu}b_{\nu}X^{\mu}X^{\nu}$ is gauged away. In order to verify the degrees of freedom of these two cases, we are going to deal with the Lagrangian expressions (\ref{Superficie1}) and (\ref{Superficie2}), and use the analogous method of the matter-gauge case. So we now have two operators: ${\cal O}(W)$, ${\cal O}(Z)$ and their respective inverses whose are straightforward obtained. The associated to ${\cal O}(W)^{-1}$ propagator is obtained as,
\begin{eqnarray}
\label{propwp}
<W_{\mu},W_{\nu}>&=&\frac i{(k^2-\bar{\alpha}^2)}\omega_{\mu\nu}+
\frac{i(k^2-\beta^2)}{(k^2-\bar\mu^2_{+})(k^2-\bar{\mu}^2_{-})}\,
\theta_{\mu\nu},\nonumber \\
<W_{\mu},P_{\nu\lambda}>&=& <P_{\mu\nu},W_{\lambda}> \,= \frac{i
m}{(k^2-\bar\mu^2_{+})(k^2-\bar\mu^2_{-})}\,s_{\mu\nu\lambda}, \\
<P_{\mu\nu},P_{\lambda\rho}>&=&\frac i{(k^2-\bar{\alpha}^2)}
\omega_{\mu\nu,\lambda\rho}+\frac{i(k^2-\beta^2)}
{(k^2-\bar\mu^2_{+})(k^2-\bar\mu^2_{-})}\,
\theta_{\mu\nu,\lambda\rho}, \nonumber
\end{eqnarray}
where we have defined
\begin{equation}
\label{barmu} \bar\mu^2_{\pm}=\frac{\bar\alpha^2+\beta^2+2m^2\pm
\sqrt{(\bar\alpha^2+\beta^2+2m^2)^2-4\bar\alpha^2\beta^2}}{2}
\end{equation}
and $\bar{\alpha}= 4\lambda b^2$. In same way, for the perpendicular wave operator ${\cal
O}(Z)^{-1}$ result in,
\begin{eqnarray}
\label{propzp}
<Z_{\mu},Z_{\nu}>&=&\frac {i}{k^2}\omega_{\mu\nu}+
\frac{i(k^2-\beta^2)}{k^2(k^2-\xi^2)} \theta_{\mu\nu},\nonumber
\\ <Z_{\mu},P_{\nu\lambda}>&=& <P_{\mu\nu},Z_{\lambda}>\,= \frac{i
m}{k^2(k^2-\xi^2)}s_{\mu\nu\lambda},
\\
<P_{\mu\nu},P_{\lambda\rho}>&=&\frac i{(k^2-\beta^2)}
\omega_{\mu\nu,\lambda\rho}+\frac{i}
{(k^2-\xi^2)}\theta_{\mu\nu,\lambda\rho},\nonumber
\end{eqnarray}
where $\xi^2= \beta^2+ 2m^2$.
In order to guarantee  unitarity,  we must have the real part of the current-current amplitude greater than zero.  We can observe from the propagators (\ref{propwp}), that the longitudinal sectors of the fields $W$ and $P$ exhibit no tachyons and they are not dynamical. On the other hand, the transverse degrees of freedom of the non-mixing terms are physical as far as we assume in the expression (\ref{barmu}) that $\bar\mu^2_{+}>\bar\mu^2_{-}$, $\bar\mu^2_{+}>\bar\beta^2$ and $\bar\alpha^2>\bar\mu^2_{-}$. As a consequence, the ${\cal L}(W)$ has a pole $k^2=\bar\mu^2_{+}$ for the propagator $<W_{\mu},W_{\nu}>^T$, and a pole
$k^2=\bar\mu^2_{-}$ for the propagator $<W_{\mu\nu},W_{\lambda\rho}>^T$; they  are dynamical physical excitations.

\section{Conclusions}

In this paper, we have presented a charged vector-tensor (CSKR) matter field model which shows a connection between vector (1-form) field $B_{\mu}$ and an antisymmetric tensor (2-form) field $H_{\mu\nu}$ via a topological interaction in $(1+3)D$. Furthermore, it presents a global $U(1)$ symmetry. We have shown that the introduction of a self-interacting $B^{4}$-type potential in Lagrangian (\ref{Lag1}) is necessary to  define a topological 2-form (tensor) current (\ref{topological}). Thus, we propose a vector-tensor current which is a doublet where it accommodates the ordinary Noether current $J_{\mu}$ along with the topological 2-form current $J_{\mu\nu}$. With this procedure, we can obtain the physical spectrum of the model in a direct way, where we verify the existence of two distinct simple physical (non-tachyonic and non-ghost) poles with masses $\mu_{+}$ and $\mu_{-}$ for the transverse sectors. We emphasize that, in spite of the peculiar form of the model, the longitudinal ones decouples. The propagator poles include topological ($\alpha^{2}$) and Proca ($\beta^{2}$) mass parameters what implies that the very same physical degrees of freedom are present in the propagators of the 1-form  $B_{\mu}$ (KR) field and the 2-form $H_{\mu\nu}$ (KB) field. 
In the charged case, we introduce a gauge interacting field $A_{\mu}$ and we explore the low-energy dynamics of the model, by studying the spontaneous symmetry breaking (SSB) mechanism. The quartic form of the potential energy of the vector-tensor matter field indicates the SSB mechanism could take place for this field. Only the vector (1-form) $B_{\mu}$ (KR) field can reach the very minimum of energy of the model, consequently it is responsible for SSB mechanism. On the other hand, the vacuum energy value of this field  naturally violates the Lorentz symmetry which fixed a vector deviation to the minimum and implies contributions to the splitting of the mass spectrum. We observe that the possibilities of the fixed vector has to be of physical consistency preventing ghost and tachyon degrees of freedom, and so following in an analogous way the classification obtained in Ref. \cite{belich}. The effects of Lorentz-violating terms have recently been an object of study due to astrophysical effects and to condensed matter \cite{astro,cond}. We have studied the contributions of the fixed vector $b_{\mu}$ to the mass at tree-level, where to this purpose we suitable introduce the unitary gauge and we redefine the fields and parameters in this effective resulting model. We remark that the field $Y_{\mu}$ is gauged away and we consequently recover the usual gauge transformation of the dynamical elements of the model. We obtain two independent dynamical systems: one describes a model where a KB-field interacts with a gauge field; and another a neutral KB-field interacting with the ``Higgs'' vector field. We emphasize that this model presents a new neutral vector particle without the introduction of an ``extra'' $U(1)$ as it has been largely explored in the literature. Indeed, many phenomenological works have been proposed with the aim of suggesting and extra symmetry to account for discrepancies in the Standard Model, particularly the $Z-Z'$ mixing \cite{extra}. In astrophysical models coming from high energy considerations, it has been suggested that non-baryonic dark matter could have exotic astrophysical origin where a possible mirror matter described by extra symmetries \cite{mirror1,mirror} emerges as an interesting possibility. In our case, an extra and/or mirror matter has a topological origin as it was suggested in the Ref. \cite{mirror1} and the Lorentz violation could imply optical astrophysical effects that influence the red-shift deviations, the anisotropy of the Cosmic Radiation Background, and possibly masks the observational effects of topological defects.       
Finally, we analyze the spectrum of the two independent dynamical models obtained after the SSB mechanism. In the gauge-KR sector, we obtain the condition on the mass parameters and the fixed vector so has to have a physically consistent condition where the mass term to the 2-form matter field is crucial. We obtain that the longitudinal part has no dynamics. The result is that this sector could represent the effective dynamics of a charged spin-1 particle. On the other hand, the Higgs-KR sector can represent the dynamics of a massive neutral particle that, due to the Lorentz violation, can only be analyzed in the longitudinal and transverse projected degrees of freedom. Indeed, from another way this particular feature of Lorentz violation is emphasized in its classical electromagnetic version\cite{viol-em}, and in the perturbation studies of the model\cite{viol-pert}. We obtain the conditions on the mass parameters (topological and not) to get physically consistent degrees of freedom at tree-level. In this sector, the fixed vector $b_{\mu}$ dictates a preferred direction in space. We can conclude to given the perspective to compute a dimensional reduction of the KR model from $(1+3)$D to $(1+2)$D, where we can conjecture the existence of charged ``vector-solitons'' derived off topological solutions in $(1+2)$D, which is the study of a forthcoming work \cite{vecsoliton}.

\section{Acknowledgments}
\noindent
The authors would like to thank J.A. Helay\"el-Neto for the suggestion of the subject and discussions. A.L.M.A Nogueira is acknowledged for comments and suggestions. LPC would like CCP-CBPF for the kind hospitality. ALAP and WCS are grateful to CAPES and CNPq respectively for their Graduate fellowships.

\end{document}